\documentclass[sigconf]{acmart}
\AtBeginDocument{%
  }

\copyrightyear{2026}
\acmYear{2026}
\setcopyright{cc}
\setcctype{by}
\acmConference[WSESE '26]{3rd International Workshop on Methodological Issues with Empirical Studies in Software Engineering }{April 12--18, 2026}{Rio de Janeiro, Brazil}
\acmBooktitle{3rd International Workshop on Methodological Issues with Empirical Studies in Software Engineering (WSESE '26), April 12--18, 2026, Rio de Janeiro, Brazil}
\acmPrice{}
\acmDOI{10.1145/3786149.3788298}
\acmISBN{979-8-4007-2382-7/2026/04}
\newcommand{\lessonbox}[2]{%
\begin{center}
\fbox{%
\parbox{0.45\textwidth}{%
\vspace{2pt}%
\textbf{#1}\par
\vspace{2pt}\hrule\vspace{4pt}%
#2\par
\vspace{2pt}%
}%
}%
\end{center}
}
\begin{document}

\title[Large Language Model to Support a Systematic Mapping Study: A Practitioner's View]{On the Use of a Large Language Model to Support the Conduction of a Systematic Mapping Study: A Brief Report from a Practitioner's View}

\author{Cauã Ferreira Barros}
%%\authornote{Both authors contributed equally to this research.}
%%\orcid{0009-0002-3621-7853}
%%\authornotemark[1]
\affiliation{%
  \institution{Federal University of Goiás\\Informatics Institute}
  \city{Goiânia}
  \state{Goiás}
  \country{Brazil}}
\email{cauabarros@ufg.br}

\author{Marcos Kalinowski}
\affiliation{%
  \institution{Pontifical Catholic University of Rio de Janeiro\\Department of Informatics}
  \city{Rio de Janeiro}
  \state{Rio de Janeiro}
  \country{Brazil}}
\email{kalinowski@inf.puc-rio.br}

\author{Mohamad Kassab}
\affiliation{%
  \institution{Boston University\\Department of Computer Science}
  \city{Boston}
  \state{Massachusetts}
  \country{USA}}
\email{mkassab@bu.edu}

\author{Valdemar V. Graciano Neto}
\affiliation{%
  \institution{Federal University of Goiás\\Informatics Institute}
  \city{Goiânia}
  \state{Goiás}
  \country{Brazil}}
\email{valdemarneto@ufg.br}

\renewcommand{\shortauthors}{Barros et al.}

\begin{abstract}
The use of Large Language Models (LLMs) has drawn growing interest within the scientific community. LLMs can handle large volumes of textual data and support methods for evidence synthesis. Although recent studies highlight the potential of LLMs to accelerate screening and data extraction steps in systematic reviews, detailed reports of their practical application throughout the entire process remain scarce. This paper presents an experience report on the conduction of a systematic mapping study with the support of LLMs, describing the steps followed, the necessary adjustments, and the main challenges faced. Positive aspects are discussed, such as (i) the significant reduction of time in repetitive tasks and (ii) greater standardization in data extraction, as well as negative aspects, including (i) considerable effort to build reliable well-structured prompts, especially for less experienced users, since achieving effective prompts may require several iterations and testing, which can partially offset the expected time savings, (ii) the occurrence of hallucinations, and (iii) the need for constant manual verification. As a contribution, this work offers lessons learned and practical recommendations for researchers interested in adopting LLMs in systematic mappings and reviews, highlighting both efficiency gains and methodological risks and limitations to be considered.
\end{abstract}



%%  \includegraphics[width=\linewidth]{img/imagem-artigo.pdf}
%%  \caption{Teaser of the studie.}
%%  \Description{It summarizes the main point of the article, reducing the time taken to perform tasks.}


\maketitle

\section{Introduction}

Systematic mappings and reviews can be considered rigorous methods for consolidating scientific knowledge, especially in the field of Software Engineering. However, the conduction of such efforts involves long and laborious steps, such as screening a large number of studies, carefully reading full texts, and performing structured data extraction. These processes, in addition to requiring weeks or even months of dedication, are also subject to inconsistencies among reviewers and the impact of cognitive fatigue \cite{Kitchenham2007, Wohlin2022}.

In this context, Large Language Models (LLMs) emerge as alternatives capable of supporting the repetitive and intensive stages of these methods. Recent studies have demonstrated that LLMs (i) can accelerate the initial screening of articles, (ii) assist in extracting relevant information, and even (iii) improve consistency among reviewers \cite{Syriani2024, Polak2024}. Despite these advances, the literature still focuses on feasibility analyses or experiments restricted to specific stages, leaving open questions regarding the practical application of these models throughout the entire cycle of a systematic review or mapping \cite{Bano2024, DePaoli2024}.

The systematic mapping study discussed in this experience report originally investigated the use of Large Language Models to support qualitative research and qualitative analysis tasks. This mapping aimed at identifying how LLMs have been applied in qualitative studies, as well as their reported benefits and challenges. In the present work, we revisit the same set of primary studies and focus on reporting the experience of automating and comparing key stages of that mapping—specifically study selection and data extraction—using LLMs. Therefore, this article does not introduce new domain findings, but rather analyzes the practical implications of LLM-supported execution in an existing systematic mapping study (SMS).

The main contribution of this paper is an experience report on the execution of a systematic mapping study (SMS) with the support of LLMs. It describes the points where the application proved to be advantageous, the challenges faced during execution, and the adjustments necessary to ensure accuracy. The report aims to provide practical contributions to the community by offering time comparisons, accuracy analyses, and critical reflections on risks and verification, as well as proposing recommendations for researchers intending to integrate LLMs into their own systematic studies.

\section{Background}

LLMs are artificial intelligence architectures trained on large volumes of textual data with the goal of recognizing complex patterns and generating linguistic outputs similar to human production \cite{Min2023, Kumar2024}. Models such as ChatGPT\footnote{\url{https://chat.openai.com}}, LLaMA\footnote{\url{https://ai.meta.com/llama/}}, Copilot\footnote{\url{https://copilot.microsoft.com/}}, Claude\footnote{\url{https://claude.ai/}}, and Gemini\footnote{\url{https://gemini.google.com}} have been applied to tasks ranging from machine translation and text summarization to sentiment analysis and programming support \cite{Barros2025}. %In the field of qualitative research, for example, LLMs have gained prominence due to their ability to assist with intensive tasks such as initial data coding, categorization, and theme extraction \cite{Ray2023, Bano2024}.

%Traditional qualitative analysis, grounded in methods such as Grounded Theory \cite{Corbin2014, Charmaz2006}, is characterized by being time-intensive and susceptible to researcher subjectivity. Processes of reading, coding, and synthesizing data often require weeks or months, especially when dealing with large volumes of interviews or digital records. In this scenario, 
LLMs offer the potential to accelerate time to fulfill some research stages, reduce individual biases, and enhance the scalability of analysis, as demonstrated in applications in the health domain \cite{Mathis2024} and education \cite{Syriani2024}. Nevertheless, the literature recognizes that the use of these models does not replace human interpretation but rather serves as complementary support for more consistent and reproducible analyses \cite{Bano2024}.

In the specific context of Software Engineering, systematic reviews and mappings are characterized by their structured and evidence-based procedures, designed to ensure methodological rigor and transparency in knowledge synthesis \cite{Kitchenham2007, Wohlin2022}. However, these methods face challenges related to large-scale screening and data extraction, which opens space for the adoption of LLMs. Recent studies, such as those by Syriani et al. \cite{Syriani2024} and Polak and Morgan \cite{Polak2024}, have already pointed out the feasibility of employing models like ChatGPT in critical stages of systematic reviews, presenting gains in efficiency and accuracy. Other studies, however, highlight risks such as dependence on prompt engineering, inconsistency across model versions, and the possibility of hallucinations \cite{DePaoli2024}. Therefore, there is a clear need to explore empirical reports that evaluate the use of LLMs in systematic mappings, identifying effective practices and remaining limitations.
\\\\
\textbf{Related Work}. Several studies have investigated the use of LLMs to support screening and data extraction in systematic reviews. \citet{Khraisha2024} showed that GPT-4 can achieve performance comparable to human reviewers under well-structured prompts, but highlighted strong sensitivity to prompt design and data imbalance. Similarly, \citet{DelgadoChaves2025} evaluated 18 LLMs across biomedical reviews and reported workload reductions ranging from 33\% to 93\%, while emphasizing that performance depends heavily on task definition and inclusion/exclusion clarity, requiring human oversight.

Focusing on domain-specific applications, \citet{LopezPineda2025} assessed LLaMA-3 and ChatGPT-4o mini for title and abstract screening in biomedical reviews, reporting high precision and reduced human effort, though limited to a single domain. In Software Engineering, \citet{Huotala2024} compared GPT-3.5 and GPT-4 using different prompting strategies and found that GPT-4 and Few-shot prompting improved accuracy, but full automation remained infeasible without expert supervision.

Beyond screening performance, reproducibility and methodological transparency have been identified as key challenges. \citet{Felizardo2025} highlighted inconsistencies across model versions and insufficient prompt reporting as major threats to replicability. Complementing this perspective, \citet{Chen2025} compared fully automated and semi-automated LLM-based selection pipelines, showing that semi-automated approaches achieved higher accuracy (82.7\% correct inclusions and 92.2\% correct exclusions), reinforcing that human-in-the-loop strategies currently offer the most reliable balance between efficiency and rigor.
\\\\
\noindent
\textbf{\underline{Our contribution compared to the related work.}}\\ Although these works demonstrate important advances, they mainly address specific aspects of LLM usage in reviews (screening, data extraction, or replication challenges), but they do not present an integrated account of the end-to-end experience of conducting an SMS. None of them provide detailed comparisons of execution time with and without LLMs, quantitative accuracy assessments across all stages, reports of errors encountered during practice, or descriptions of the real-time adjustments required. The present study seeks to fill this gap by providing a complete empirical account, combining quantitative and qualitative evidence on the use of LLMs in a systematic mapping study, in addition to proposing practical recommendations for researchers interested in applying this approach.

\section{Method and Experience Conduction}
This section presents the methodological approach adopted for conducting the SMS with LLM support. 
\\\\
\noindent \textbf{Disclaimer.} We need to clarify that when we mention the \textit{manual} extraction in a first step of this study, this actually refers to the manual extraction performed by us and reported in a prior study of ours \cite{Barros2025}. Now, we revisited the same studies of prior published work, since we are extending it; then, the same studies were extracted, but now using LLM. The same 219 studies used for first selection in \citet{Barros2025} was also used for screening with LLM. The SMS is not published, yet; then, only its conduction is reported here. The SMS is an extension of \citet{Barros2025}. The manual procedures reported in \citet{Barros2025} were conducted between November 2024 and January 2025, whereas the new LLM-based conduction occurred between July and October 2025. Because of this temporal gap, the reported estimates of manual execution time may not be perfectly precise, as some degree of recall bias or loss of reference regarding the exact duration of manual activities may have occurred.
\\\\
\noindent \textbf{Protocol Definition.} 
%\subsection{Protocol Definition}
The starting point was the development of a protocol following the guidelines of Kitchenham and Charters \cite{Kitchenham2007} and Wohlin et al. \cite{Wohlin2022}. The research questions, the databases to be consulted, the inclusion and exclusion criteria, and the data extraction procedure were defined. This protocol was essential to ensure methodological rigor and comparability of results, serving as a guide for both the manual application and the application with LLM support.

\noindent \textbf{Search and Selection Strategy}: The searches were carried out in widely recognized digital databases (Scopus\footnote{https://www.scopus.com/}, IEEE Xplore\footnote{https://ieeexplore.ieee.org/}, ACM Digital Library \footnote{https://dl.acm.org/}, among others). In a first moment, the screening of titles and abstracts was conducted manually, following the established criteria. Subsequently, the same process was performed with the support of ChatGPT-4, using a carefully structured prompt that explicitly stated the inclusion and exclusion criteria. This approach made it possible to compare the efficiency of the manual and automated processes, identifying both gains and limitations.

\noindent \textbf{Data Extraction}: The data extraction stage was conducted under two fully executed conditions across the entire study scope: (i) manually, using standardized forms completed by human reviewers, and (ii) automatically, with LLM support, using extraction templates in questionnaire format. For each selected article, both procedures were independently performed in full, allowing direct comparison between manual and automated results. The model outputs were then manually reviewed to validate consistency and accuracy.

\noindent \textbf{Verification and Risk Mitigation}: Acknowledging the risks of hallucination and bias in LLMs, a double-checking strategy was adopted. First, each piece of data automatically extracted was reviewed by at least one human reviewer. Second, discrepancies between the two approaches (manual and automated) were recorded and discussed after full-text reading, in order to understand the limits of the model’s contribution. This verification was essential to mitigate factual errors and inconsistencies.

\section{Experience Report}

\noindent
This section reports the empirical results of the LLM-supported SMS execution, focusing on time, accuracy, prompt adjustments, and exploratory tests with other LLMs.
%This section presents the experience of conducting a complete SMS with the support of LLMs. The goal is to provide an empirical account of how these models performed in practical conditions. The report describes the main stages carried out — screening, data extraction, and model testing — emphasizing the methodological adjustments, validation strategies, and quantitative comparisons that guided the evaluation. %By documenting this process, we aim to offer insights that can inform future studies seeking to integrate LLMs into evidence synthesis workflows.

\subsection{Comparison of Time and Accuracy}
One of the central objectives was to compare the performance between the two methods. The complete manual process of screening and data extraction required approximately 23 days for the screening of 219 studies and 7 days for the data extraction of 13 studies, totaling around 30 days of work conducted by two researchers \cite{Barros2025}. In contrast, the process supported by the LLM was completed in approximately 9 hours for screening the 219 studies and 1 hour for extracting data from 13 studies, representing a significant reduction in the required time. It is important to note that the manual execution did not occur continuously over the 30 days, as it would be unfeasible to maintain this activity without breaks. Table~\ref{tab:manual_vs_llm} summarizes the comparison between the manual and LLM-assisted execution, highlighting the time savings, accuracy rates, and verification approaches adopted.  

\begin{table}[htbp]
  \caption{Comparison between Manual Execution and LLM-Assisted Execution.}
  \label{tab:manual_vs_llm}
  \begin{tabular}{|p{1.5cm}|p{2.5cm}|p{3cm}|}
    \toprule
    \textbf{Aspect} & \textbf{Manual Execution} & \textbf{LLM-Assisted Execution (ChatGPT-4)} \\
    \midrule
    Screening Time & Approximately 23 days (219 studies) & Approximately 9 hours (reduction of ~98\%) \\
    \hline
    Extraction Time & Approximately 7 days (13 studies) & Approximately 1 hour (reduction of ~99\%) \\
    \hline
    Screening Accuracy & Defined by consensus among reviewers & Approximately 95\% agreement (208 correct out of 219 studies; 11 hallucinations identified) \\
    \hline
    Extraction Accuracy & High, but subject to inconsistencies among reviewers & Approximately 92.3\% agreement (12 correct out of 13 studies; 1 error identified) \\
    \hline
    Main Risks & Human reading errors or fatigue & Hallucinations, dependence on prompt engineering, inconsistency across model versions \\
    \hline
    Verification Applied & Cross-checking among human reviewers & Double-checking: comparison with manual results + review of discrepancies \\
    \bottomrule
  \end{tabular}
\end{table}

Regarding accuracy, the agreement between human reviewers was considered the reference standard. In the screening stage (title and abstract reading), the LLM achieved approximately 95.0\% accuracy (208 correct out of 219 studies), with 11 hallucination cases in which it began answering the questionnaire with information from other works. In the data extraction stage, the LLM reached approximately 92\% accuracy (12 correct out of 13 studies), with 1 error identified. These results indicate the potential to accelerate repetitive stages but remain insufficient to eliminate the need for rigorous human supervision, especially to detect hallucinations and inconsistencies.

In this report, LLM-supported activities refer only to executing study selection and data extraction using previously defined prompts. Prompt design and iterative refinement were performed prior to execution and were not included in the reported LLM time.

\subsection{Adjustments During Conduction}
Throughout the screening and extraction processes, it was necessary to adjust the prompts four times and six times, respectively, mainly to address ambiguities and improve the standardization of outputs. Furthermore, some stages, such as the analysis of the extracted data, proved to be unfeasible to automate reliably, since in more than half of the cases the model retrieved information from sources other than the target study. This reinforces the importance of human judgment in ensuring the accurate analysis of the data obtained.

\subsection{Tests with Other Models}
With the aim of comparing performance and identifying promising LLMs, additional tests were conducted after validating the outputs obtained by ChatGPT. The models Gemini PRO, Manus, and Copilot were evaluated using subsets of studies previously classified correctly by ChatGPT. In the screening process, 50 studies were analyzed: Manus correctly classified 49 (98\%), Gemini PRO 45 (90\%), and Copilot 30 (60\%). In the data extraction process, 10 studies were considered: Manus achieved 4 correct (40\%), Gemini PRO 9 (90\%), and Copilot 6 (60\%). Table~\ref{tab:other_llms} presents the performance results for each model in both the screening and data extraction tasks.

\begin{table}[htbp]
  \caption{Accuracy of other LLM models.}
  \label{tab:other_llms}
  \begin{tabular}{|p{2cm}|p{2.5cm}|p{2.5cm}|}
    \toprule
    \textbf{Model} & \textbf{Screening (50 studies)} & \textbf{Extraction (10 studies)} \\
    \midrule
    Manus      & 49 (98\%) & 4 (40\%) \\
    \hline
    Gemini PRO & 45 (90\%) & 9 (90\%) \\
    \hline
    Copilot    & 30 (60\%) & 6 (60\%)\\
    \bottomrule
  \end{tabular}
\end{table}

The purpose of this work, and particularly of this section, is not to compare other models to determine whether their performance is superior or inferior to ChatGPT. Rather, the objective is to explore models with potential for future investigation. That said, the results reveal a significant variation in performance among the models, with Gemini PRO standing out for its robustness in data extraction and Manus in screening, while Copilot showed inferior performance in both scenarios.

\section{Discussion}

This section discusses the empirical findings reported in the previous section, focusing on their implications for the use of LLMs in systematic mapping studies.

\subsection{Model Variability and Task Dependency}

The comparative evaluation of different LLMs revealed substantial variability in performance across tasks. Gemini PRO achieved stronger results in data extraction, Manus performed best in the screening stage, and Copilot showed inferior performance in both tasks. These results indicate that LLM effectiveness is task-dependent and cannot be assumed to be consistent across different stages of a systematic mapping study.

As shown in Table \ref{tab:other_llms}, the models exhibited task-dependent performance, reinforcing that LLM adoption should be preceded by task-specific empirical validation.

Rather than identifying a universally superior model, these findings emphasize that model selection should be treated as a methodological decision, grounded in empirical validation and aligned with the specific objectives of each stage of the mapping process.

\subsection{Sample Size and Scope Limitations}

This experience report is subject to limitations related to sample size and scope. While the screening stage involved 219 studies, the data extraction phase was conducted on a smaller subset of 13 studies. In addition, the evaluation of other LLMs relied on reduced samples (50 studies for screening and 10 for extraction), which constrains the generalizability of the observed results.

These sample sizes are sufficient for an experience report focused on practical feasibility, observed behavior, and methodological implications. However, they do not support strong statistical claims or broad generalizations regarding model performance. Future studies with larger datasets and multiple domains are required to validate the trends identified in this work.

\subsection{Interpretive Nature of SMS and Implications for LLM Support}

Systematic mapping studies in software engineering combine interpretive activities with operational and repetitive tasks. While judgments about relevance, meaning, and synthesis require contextual understanding and critical reading, other stages—such as large-scale screening and structured data extraction—involve substantial mechanical effort. In this context, the time reductions reported in this study should not be interpreted as a proxy for methodological quality, but rather as evidence that LLMs can selectively support low-level and repetitive activities without replacing human interpretation.

Nevertheless, reliance on LLM outputs—even when followed by human verification—introduces important risks. Verification presupposes that researchers possess sufficient familiarity with the primary studies to recognize omissions, misinterpretations, or oversimplifications. If engagement is limited to model-generated outputs, relevant information that was not extracted may remain unnoticed, potentially weakening later interpretive stages. This concern is particularly relevant from a training perspective, as systematic mappings also function as a learning process through which researchers develop domain understanding and methodological rigor.

Taken together, these observations reinforce that LLMs should be integrated cautiously and deliberately, as complementary tools within human-centered workflows. Their value lies in supporting efficiency in operational stages, while preserving direct engagement with the source material as a prerequisite for trustworthy interpretation and synthesis.

\section{Lessons Learned}
Based on the reported experience, some recommendations can guide researchers interested in adopting LLMs in systematic reviews and mappings. First, it is essential to document prompts in detail and keep them consistent in order to ensure reproducibility. Second, a manual verification strategy should be adopted in all critical stages, especially in screening and data extraction, to mitigate hallucinations and interpretation errors. In addition, the use of structured templates for data extraction is recommended to increase the standardization of results. It also proved important to continually remind the model of the task’s objective, thereby reducing context drift and maintaining coherence in responses. Finally, it is crucial to restrict the use of LLMs to mechanical and repetitive tasks, preserving human judgment in interpretive and creative stages, thus ensuring methodological rigor without compromising efficiency.

In this study, structured templates refer to predefined extraction prompts designed as fixed questionnaires. These templates define, in advance, the exact set of questions to be answered for each primary study, ensuring that the same information is systematically extracted across all articles.
\\\\
\noindent \textbf{Take away Lessons}. The conduction of an SMS with LLM support allowed us to identify a set of relevant lessons that can be used by the research community, as follows.

\lessonbox{Lesson 1: Time Efficiency is Significant but Contextual}{LLMs can drastically reduce execution time in screening and extraction, but these gains are context-dependent and still require human supervision.}

The main advantage observed was the significant reduction in execution time during the screening and extraction stages. The LLM accelerated repetitive and low-complexity tasks, allowing researchers to focus on higher-value analytical activities. However, these efficiency gains do not eliminate the need for human supervision and should be interpreted in context. The magnitude of the observed reduction depends on factors such as task characteristics, data quality, and prompt design. In addition, the manual and LLM-assisted executions were conducted in different periods, which may introduce minor imprecision in the reported time estimates. Therefore, while the reduction in execution time is substantial, it should be understood as a context-dependent improvement rather than a universal outcome.

\textit{Evidence note.} Manual screening and extraction required approximately 23 days and 7 days respectively, whereas the LLM completed the same activities in about 9 hours and 1 hour, corresponding to a reduction of approximately 98–99\%. These estimates are indicative, as the manual (Nov 2024–Jan 2025) and LLM-assisted (Jul–Oct 2025) executions were performed in different time periods.

\lessonbox{Lesson 2: Prompt Design is Critical}{Small changes in prompts can produce large variations in accuracy.}

The dependency on well-structured prompts proved critical: small variations in formulation resulted in notable changes in response accuracy. In addition, hallucinations and inaccurate extractions occurred. This reinforces the need for manual verification and clear documentation of the prompts used, in order to ensure reproducibility.

For instance, open-ended questions tended to yield verbose answers and occasional unsupported additions, whereas option-based answers produced more consistent outputs and fewer hallucinations.

\textit{Evidence note.} During the study, the screening prompt was refined four times and the extraction prompt six times to improve precision and reduce ambiguity, demonstrating the sensitivity of LLM performance to prompt wording.

\lessonbox{Lesson 3: Human Oversight Remains Essential}{Automation supports execution, but interpretation and verification must remain under human control.}

Although LLMs proved effective in supporting repetitive execution tasks, human oversight remained essential throughout the study. Interpretive activities, such as synthesizing findings and categorizing contributions, require contextual and methodological judgment that the model could not reliably demonstrate. In addition, human verification was necessary to identify hallucinations and inconsistencies in both screening and data extraction outputs. These observations reinforce that automation can support execution, but cannot replace human responsibility in ensuring analytical rigor and factual correctness.

\textit{Evidence note.} Eleven hallucinations were identified during the screening of 219 studies, and one incorrect extraction was detected among 13 analyzed papers, confirming the need for continuous human verification and interpretive control.

\lessonbox{Lesson 4: Hybrid Workflows are the Way Forward}{LLMs are valuable support tools, but not self-sufficient replacements.}

From a scientific perspective, the experience revealed that the use of LLMs in systematic mappings is feasible but not self-sufficient. These models can serve as valuable support tools, provided they are embedded in workflows that combine partial automation with targeted human supervision. Complete delegation to the model still entails risks of losing methodological rigor.

\textit{Evidence note.} Accuracy levels of approximately 95\% in screening and 92\% in data extraction were achieved only after manual verification, indicating that a hybrid human-in-the-loop workflow was necessary to ensure reliable results.

\lessonbox{Lesson 5: Gemini PRO Shows Promise}{Gemini PRO demonstrated robust accuracy and deserves attention in future research.}

Gemini PRO showed promising results in the tests with other models, approaching the performance of ChatGPT and surpassing other models tested. This indicates that Gemini PRO is a model worth monitoring in future research, as it may represent a viable alternative for systematic mappings and reviews.

\textit{Evidence note.} In comparative tests, Gemini PRO achieved 90\% accuracy in data extraction (9 correct out of 10 studies) and 90\% accuracy in screening (45 correct out of 50 studies), outperforming Copilot and closely matching the performance observed with ChatGPT.

\section{Limitations and Threats to Validity}
This experience report has limitations regarding generalizability. The procedures, challenges, and outcomes reflect the experience of a single researcher conducting a systematic mapping study with LLM support. Results may vary for other researchers depending on factors such as expertise, application domain, prompt design, and interpretation strategy. Therefore, the findings should be interpreted as context-dependent rather than universal.

A second limitation concerns the comparison of execution time between manual and LLM-assisted processes. The manual procedure was conducted first, followed by the LLM-assisted execution, which extended the overall duration of the study. This sequential execution may affect the reported time estimates, as parallel or controlled executions could lead to different results. Thus, the time comparisons should be considered indicative rather than absolute.

Despite these limitations, the study provides empirical evidence on the practical use of LLMs in systematic mappings and offers insights to support future research and methodological refinement.

\section{Conclusion}
This experience report has shown that the use of LLMs in conducting a systematic mapping study can yield significant gains in terms of efficiency and initial consistency, particularly in the screening and data extraction stages. At the same time, it demonstrated that these gains are accompanied by risks and limitations that cannot be overlooked, such as dependence on prompts, the occurrence of hallucinations, and the need for continuous human verification.

Unlike related works that evaluated isolated stages or discussed general challenges, this study provided an integrated end-to-end account, presenting comparisons of time, accuracy, and necessary adjustments during practical conduction. As a scientific contribution, the article demonstrates that LLMs are promising tools to support systematic reviews, but they must be embedded in hybrid workflows, where partial automation and critical supervision go hand in hand.

Thus, the main message is clear: LLMs do not replace the role of researchers in systematic mappings, but they can enhance execution capacity and accelerate processes when used judiciously. Future progress will depend not only on technological improvements in the models but also on the development of solid methodological protocols that guide their responsible integration into empirical studies.

\begin{acks}
ChatGPT was also employed for text translation into English, with all versions subsequently revised manually to ensure precision, clarity, and consistency. All translations were further verified by human reviewers regarding spelling, grammar, and methodological integrity. This study was financed in part by the Coordenação de Aperfeiçoamento de Pessoal de Nível Superior – Brazil (\textbf{CAPES}) – Finance Code 001. This work has also been partially funded by the project Research and Development of Computational Techniques for Security and Privacy of Second-Generation Multimodal Data, supported by the Advanced Knowledge Center in Immersive Technologies (\textbf{AKCIT}), with financial resources from the PPI IoT of the MCTI grant number 057/2023, signed with \textbf{EMBRAPII}. The support provided by these funding sources was fundamental for both the execution and publication of this research.

\paragraph{Data Availability Statement.}
All prompts and LLM outputs used in this study are publicly available on Zenodo for reproducibility purposes at: \url{https://doi.org/10.5281/zenodo.14177022}.
\end{acks}

\bibliographystyle{ACM-Reference-Format}
\bibliography{sample-base}

%\appendix

%\section{Tables}

\end{document}